\documentstyle[aps,float,epsf]{revtex}    
\newcommand{\be}{\begin{equation}}
\newcommand{\ee}{\end{equation}}
\newcommand{\half}{\frac{1}{2}}
\newcommand{\bea}{\begin{eqnarray}}
\newcommand{\eea}{\end{eqnarray}}
\newcommand{\nn}{\nonumber}

\newcommand{\k}{\vec{k}}
\newcommand{\intk}{\int \frac{d^3k}{(2\pi)^3}}
\newcommand{\intp}{\int \frac{d^3p}{(2\pi)^3}}
\newcommand{\D}{\mbox{D}}
\begin{document}
\preprint{ITFA-99-13}
\title{Counterterms for Linear Divergences in Real-Time 
Classical Gauge Theories\\ at High Temperature }
\author{B. J. Nauta\footnote{E-mail: nauta@wins.uva.nl}}
\address{Institute for Theoretical Physics, University of Amsterdam\\
        Valckenierstraat 65, 1018 XE Amsterdam, The Netherlands}
\maketitle
\vspace{1cm}
\begin{abstract}
Real-time classical SU($N$) gauge theories at non-zero temperature
contain linear divergences.  
We introduce counterterms for these divergences in 
the equations of motion in the continuum and on the lattice. 
These counterterms can be given in terms of 
auxiliary fields that satisfy local 
equations of motion. 
We present a lattice model with 6+1D auxiliary fields that for IR-sensitive 
quantities yields cut-off independent results to leading order in the coupling.
Also an approximation with 5+1D auxiliary fields is discussed.
\end{abstract}
\section{Introduction}\label{sec1}
In recent years there has been a considerable interest in the calculation
of the Chern-Simons diffusion rate in the symmetric phase of the electroweak 
theory.
By the chiral anomaly this diffusion rate is related to the rate of 
baryon number 
non-conservation \cite{hooft}
and therefore it is an important parameter in electroweak
baryogenesis scenarios; see for  reviews e.g. \cite{rubakov,trodden}.
It was originally suggested by Grigoriev and Rubakov \cite{grigoriev}
that such infrared sensitive quantities 
could be determined from a 
classical theory. 
Their idea was that field configurations 
at momenta \(k<<T\)  are expected to behave classically:  
\be
n(k)=\frac{1}{e^{\beta k}-1}\sim 
\frac{1}{\beta k}=n_{\rm cl}(k)\;.
\ee
Since the diffusion of Chern-Simons number is dominated by soft fields with 
typical momenta \(k\sim g^2T\), it was expected that it could be calculated 
in a classical theory \cite{ambjorn,smit}.
Later, it was argued by Arnold, Son and Yaffe \cite{asy} that this picture 
is not complete, because soft field 
configurations are sensitive to (particle-like) thermal fluctuations with 
hard momenta \(k\sim T\), that are not correctly reproduced by the classical 
theory.  
These hard modes introduce Landau damping and slow down the dynamics of 
nearly static magnetic fields at momenta \(k\sim g^2T\).
Including these hard modes leads to an estimate of
a typical time-scale \(t\sim (g^4T)^{-1}\) of  these fields.
B\"{o}deker \cite{bodeker}
showed that also semi-hard modes at momentum scale \(gT\) 
contribute to leading order, and a even logarithmic correction 
\(t\sim (g^4T\log\frac{1}{g})^{-1}\)
to the time-scale is obtained.
This logarithmic correction has been calculated in various approaches
\cite{asy2,asy3,litim,blazoit2,valle,bodeker2,bodeker3}.

To calculate the Chern-Simons diffusion rate 
in an (effective) classical theory with cut-off \(\Lambda\),
the corrections of 
the hard modes and semi-hard modes have to be incorporated. For
the semi-hard modes this requires that the cut-off is large enough, 
\(\Lambda>>gT\).
To include the hard modes basically two approaches can be taken. One 
approach is that one uses the classical hard modes to mimic the effect of the 
hard modes in the quantum theory\cite{arnold}. 
These classical hard modes introduce linear divergences in the theory 
\cite{asy,bms}
and if one takes the (continuum) cut-off \(\Lambda\sim T\), 
the classical hard modes 
generate the (classical) hard thermal loops and
with a proper matching this
incorporates Landau damping in the correct way. 

A different approach is to include the hard thermal loops
 in the classical theory
\cite{hu,hu2,iancu,rajantie,bmr}. This can be done in an economic way by including an induced 
source in the equations of motion for the fields and to solve for the dynamics 
of the induced source in the presence of the classical fields. 
In that case one also has the problem of the linear divergences introduced
by the classical hard modes. In the continuum these can be accounted for
by adjusting the strength of the induced source \cite{iancu,aarts}. 

On the lattice both these approaches have the problem that the linear 
divergences are not rotationally invariant \cite{bms}. 
Therefore, the classical hard modes cannot correctly mimic the
continuum hard thermal loop effects, like Landau damping, nor can these be
to accounted  for by adjusting the strength of the (continuum) induced source.
The lattice spacing dependence of the classical theory has been studied in 
\cite{bodekerlaine,moorerummukainen}

The aim of this paper is to investigate the inclusion of counterterms 
for the linear divergences on a lattice. Via  a continuum model 
(section \ref{sec2})
and a lattice model (section \ref{sec3})
suitable for perturbative calculations only, we arrive at two models
that may be used for non-perturbative calculations on a lattice.

In section \ref{secfoura} the first non-perturbative model is discussed, it
contains the exact counterterms for the linear divergences on a
lattice with lattice spacing \(a_L\). 
The counterterms are generated by 6+1D auxiliary fields that satisfy 
local equations of motion.
The finite finite renormalization is 
chosen such that the theory is matched to a quantum lattice theory, with 
a smaller lattice spacing \(a_S\). For very small couplings \(g\), the  
lattice spacing \(a_S\) can be chosen to be 
small compared to the inverse temperature
and continuum results can be obtained.

The second model is presented in section \ref{secfourb}, it 
contains the counterterms
in the approximation that the external momenta of a divergent diagram
are strongly space-like (\(p_0<<|\vec{p}|\)). This model requires 5+1D 
auxiliary fields and \(g\) is unrestricted.

\section{Continuum}\label{sec2}
In this section we review the HTL equations of motion in the continuum and
show how the counterterms for the linear classical divergences may be obtained.
It is convenient to use the local formulation of the HTL equation of Blaizot 
and Iancu \cite{blaizot}. They showed that the HTL equation is equivalent to 
a set of linearized kinetic equations. The first equation describes classical 
gauge fields in the presence of an induced source
\be
\left[D_{\mu},F^{\mu\nu}(x)\right]=j^{\nu}_{\rm ind}(x)\;,
\label{claseq}\ee
with the covariant derivative \(D_{\mu}=\partial_{\mu}+igA_{\mu}\) and 
field strength 
\(F^{\mu\nu}=\partial^{\mu}A^{\nu}-\partial^{\nu}A^{\mu}+ig[A^{\mu},A^{\nu}]\).
The induced current can be written as an integral over four-momentum \(K\) 
of the source density \(j^{\nu}(K,x)\)
\be
j_{\rm ind}^{\nu}(x)=g\int\frac{d^4 K}{(2\pi)^4}j^{\nu}(x,K)\;.
\label{sourcedens}\ee
The source density satisfies the equation
\be
[K_{\mu}D^{\mu},j^{\nu}(x,K)]=2gNK^{\nu}K^{\rho}F_{\rho\sigma}(x)
\partial_{K}^{\sigma}\Delta(K)\;,
\label{eqsource}\ee
with \(N\) the number of colors and \(\Delta(K)\) the free symmetric 
two-point correlation function of the fields that have been 
integrated out to obtain equations (\ref{claseq}-\ref{eqsource}).
The factor \(2\) accounts for the spin degrees of freedom.
Equations (\ref{claseq}-\ref{eqsource}) are equivalent to the HTL's 
calculated with propagator  \(\Delta(K)\).

The kinetic equations (\ref{claseq}-\ref{eqsource})
are derived under the assumptions that the theory is weakly coupled
and that \(x\) is a slowly varying variable, \(\partial<<K\sim T\).
Since we are interested in the divergences of the theory the loop momenta
can be taken of the order of the cut-off \(k\sim \Lambda\)
which is much 
larger than the fixed external momentum \(Q\sim \partial\).

In the continuum the two-point function takes the form
\be
\Delta(K)=[\Theta(K_0)-\Theta(-K_0)]\delta(K^2)N(K_{0})\;,
\label{prop}\ee
with \(N(K_{0})\) the equilibrium distribution function of the fluctuations. 
The 
\(\delta\)-function in (\ref{prop}) forces the fluctuations on-shell with dispersion relation \(K^{0}=\pm k\), \(k=|\vec{k}|\).
Inserting the propagator (\ref{prop}) in (\ref{eqsource}), the
solution for the source density can be written as
\be
j^{\nu}(x,K)=2N K^{\nu}\delta(K^2)
\left[\Theta(K_0)\delta N(x,\vec{k})-\Theta(-K_0)\delta N(x,-\vec{k})\right]\;.
\label{partdens}\ee
The function \(\delta N(x,\vec{k})\) may be interpreted as a particle 
distribution function, because it satisfies the kinetic equation
\be
\left[v^{\mu}D_{\mu},\delta N(x,\k)\right]=gv^{\rho}F_{\rho 0}N'(k)\;,
\label{kineq}\ee
with the velocity \(v^{\mu}=(1,\vec{v})\) and
\(\vec{v}=\vec{k}/k\) and \(N'(k)\) the derivative of the equilibrium 
distribution function to the energy.

The induced source is proportional to the deviation from  equilibrium of the 
density of particles in the plasma
\be
j^{\nu}_{\rm ind}(x)=g\intk v^{\nu}\delta N(x,\k)\;,\\
\label{indcurrent}\ee
with \(g\) the gauge coupling. 

Our task is to specify what we mean by the particles of the plasma. If we 
would consider QED, the particles would be the fermions and the equilibrium 
particle distribution function would be the Fermi-Dirac distribution function.
In the SU($N$) gauge theory that we consider here, the particles are  
gauge field excitations.
If the gauge fields in (\ref{claseq}) and (\ref{kineq}) would be interpreted 
as mean fields
then the particles can be identified with the thermal fluctuations
and \(N(k)\) would be the Bose-Einstein distribution function
\(n_{BE}(k)\) \cite{blaizot}. However we consider the equations 
(\ref{claseq}-\ref{kineq})
as the equations of motion of a classical statistical theory, where still a 
thermal average over the initial fields has to be taken.
The particles in this case are the quantum fluctuations with distribution 
function 
\be
N(k)=n_{BE}(k)-T/k\;, 
\label{eqdistr}\ee
as can be derived from
separating off the classical fields from the path integration and calculating 
the two-point function of the remaining fields \cite{nauta} which appears in
(\ref{eqsource}).

The equations (\ref{claseq}-\ref{kineq}) may be simplified by introducing 
the field
\be
W(x,\vec{v})=-\delta N(x,\vec{k})\left(gN'(k)\right)^{-1}\;.
\label{defW}\ee
It follows from (\ref{kineq}) 
that \(W\) is independent of the energy \(k\), assuming this is consistent with the initial data.
The kinetic equation (\ref{kineq}) in terms of this field reads
\be
\left[v^{\mu}D_{\mu},W(x,\vec{v})\right]=\vec{v}\cdot\vec{E}(x)\;.
\label{eqW}\ee
The induced current (\ref{indcurrent}) can be written as
\be
j^{\nu}_{\rm ind}(x)=-2g^2 N \int\frac{dk}{2\pi^2}k^2
N'(k)
\int\frac{d\Omega}{4\pi}v^{\nu}W(x,\vec{v})\;,
\label{indcurrentW}\ee
where the angular integration is over the direction of $\vec{v}$.
As usual the $k$-integration decouples from the angular integration 
in the HTL approximation.

Due to the subtraction in $N(k)$ (\ref{eqdistr}), the integration over $k$ 
introduces a linear divergence. This divergence acts as a counterterm
for the linear divergences in the classical theory \cite{nauta2}.
Let us argue why this is correct.
The argument used in \cite{nauta2} is one of consistency: since the full 
quantum correlation functions cannot contain classical linear divergences, 
the quantum corrections to the classical equations should provide the correct 
counterterms. Another argument is that integrating out the classical 
fluctuations around 
a given background field generates the classical induced source in the 
effective equations for the background field. Since the classical induced 
source (in the HTL approximation) is precisely what is subtracted in 
(\ref{indcurrentW}), the equations of motion for the background field do 
not contain linear divergent terms.
The logarithmic divergences of the classical theory are more complicated and 
are not considered here.

If we use dimensional regularization, the subtraction does not contribute,
since linear divergences are set equal to zero. We then have
\be
j^{\nu}_{\rm ind}(x)=3\omega_{\rm pl}^2
\int\frac{d\Omega}{4\pi}v^{\nu}W(x,\vec{v})\;,
\label{indsource}\ee
with the plasmon frequency \(\omega_{\rm pl}^2=g^2NT^2/9\).

As a different regularization scheme that makes the linear divergence 
explicit, we may restrict the subtraction to momenta \(k<\Lambda\):
\be
N_{\Lambda}(k)=n_{BE}(k)-\frac{T}{k}\Theta(\Lambda-k)\;.
\ee
The result of the $k$-integration in the induced source  yields then
\be
-2g^2 N \int\frac{dk}{2\pi^2}k^2
N'_{\Lambda}(k)=3\omega_{\rm pl}^2-\frac{2}{\pi^2}g^2N\Lambda T\;,
\label{lindsou}\ee
where we see an explicit linear divergence. The inclusion of a linear dependence 
on the cut-off in the plasmon frequency as in (\ref{lindsou}) was proposed in 
\cite{iancu,aarts}. The derivation presented here and in more detail in 
\cite{nauta2} shows how such a cut-off dependence arises in a continuum theory
as a consequence of the subtraction in the distribution function 
(\ref{eqdistr}) for the 
quantum fluctuations.

From the point of view of renormalization the situation is quite remarkable: 
there is an infinite set of linearly divergent diagrams that are all 
non-local (or momentum-dependent) but the renormalized equations of motion
can be brought in a local form and the linear divergence is introduced in 
just one parameter. Note also that the usual HTL's are just a finite 
renormalization.

\section{Perturbative renormalization on a lattice}\label{sec3}

Before turning to the HTL equations of motion, we shortly review 
the static classical theory on a lattice, as far as the linear divergences 
are concerned. The only linear divergence in the static theory occurs in the 
Debye mass. A counterterm for this divergence may be introduced
in the mass of the temporal gauge field \cite{static,statct}
\be
\delta m^2=m_D^2-m_{\rm cl}^2\;,
\label{massct}\ee
with the continuum HTL contribution
\be
m_D^2=-2g^2N\intk n^{\prime}(k)=\frac{1}{3}g^2NT^2\;,
\label{HTLmass}\ee
and the classical mass (for a simple cubic lattice with lattice spacing \(a\))
\be
m_{\rm cl}^2=-2g^2N\intp n^{\prime}(\Omega_{\vec{p}})\approx 0.51g^2NTa^{-1}\;,
\label{classmass}\ee
where \(\vec{p}\) is restricted to the first Brillouin zone 
\(|p_{i}|\leq\pi/a\) and the energy \(\Omega_{\vec{p}}\) is 
\be
\Omega_{\vec{p}}^2=\frac{4}{a^2}\left[
\sin^2\left(\frac{p_{x}a}{2}\right)
+\sin^2\left(\frac{p_{y}a}{2}\right)+
\sin^2\left(\frac{p_{z}a}{2}\right)\right]\;.
\label{disrel}\ee
The mass (\ref{classmass}) is the linear divergent contribution to the Debye 
mass on 
the lattice, its subtraction in (\ref{massct}) ensures that no linear 
divergences 
are present in the static theory with the mass counterterm  included.
The continuum HTL contribution (\ref{HTLmass}) to the counterterm mass
(\ref{massct}) provides the finite renormalization, which is chosen such
the \(a\rightarrow 0\) limit yields the continuum result.

We want to extend this approach to the real-time classical theory.
We consider again the the equation for the gauge fields (\ref{claseq}),
but with a lattice regularization; time is continuous but space is a simple 
cubic lattice with lattice spacing \(a\).
Similar to the mass counterterm (\ref{massct}) in the static case,
we want the source to contain a continuum HTL contribution with a classical 
lattice contribution subtracted. To this end, we introduce 
two particle densities \(\delta N(x,\k)\) and \(\delta N_{\rm ct}(x,\vec{p})\)
for particles with energies \(E_{\k}=|\k|\) and \(\Omega_{\vec{p}}\) 
respectively.
The idea is that the particle density \(\delta N_{\rm ct}\) generates the 
counterterms for the linear divergences and \(\delta N\) generates the 
correct finite renormalization , the ``good'' HTL contributions.
Then these particle densities satisfy the equations
\bea
\left[v^{\mu}D_{\mu},\delta N(x,\k)\right]&=&gv^{\rho}F_{\rho 0}(x)
n'_{BE}(k)\;,\label{eqnhtl}\\
\left[v_{\rm lat}^{\mu}D_{\mu},\delta N_{\rm ct}(x,\vec{p})\right]&=&
gv_{\rm lat}^{\rho}F_{\rho 0}(x)n'_{\rm cl}(\Omega_{\vec{p}})\;.
\label{eqnct}\eea
Here and in the following, we persist in using a continuum notation although 
we use a lattice regularization for the UV-divergences.
The kinetic equation for \(\delta N(x,\k)\) is eq.(\ref{kineq})
with the Bose-Einstein distribution function as equilibrium distribution 
function.
In the kinetic equation for \(\delta N_{\rm ct}(x,\vec{p})\) the velocity
on the lattice is \(v_{\rm lat}^{\mu}=(1,\vec{v}_{\rm lat})\) with 
\cite{arnold}
\be
v^{i}_{\rm lat}=\partial_{p_i}\Omega_{\vec{p}}=\frac{1}{a\Omega_{\vec{p}}}\sin(ap_i)\;,
\label{latvel}\ee
and \(|\vec{v}_{\rm lat}|\not =1\) in general.
The induced current in the classical equations of motion for the gauge fields
then read
\be
j^{\mu}_{\rm ind}(x)=2gN\int\frac{d^3k}{(2\pi)^3}v^{\nu}\delta N(x,\k)
-2gN\int\frac{d^3 p}{(2\pi)^3}v_{\rm lat}^{\nu}\delta N_{\rm ct}(x,\vec{p})\;.
\label{latindcur}\ee
Here the integration over \(\vec{p}\) is 
restricted to the first Brillouin zone \(|p_i|<\pi/a\).

As in the continuum, a field \(W(x,\vec{v})\) may be defined that satisfies 
(the lattice version of) equation (\ref{eqW}). Since the lattice velocity 
(\ref{latvel}) is not restricted to the speed of light, we have to allow for general velocities \(\vec{v}\) in (\ref{eqW}).
The induced current (\ref{latindcur}) reads
\be
j^{\nu}_{\rm ind}(x)=3\omega_{\rm pl}^2
\int\frac{d\Omega}{4\pi}v^{\nu}W(x,\vec{v})-
2g^2NTa^{-1}\int\frac{d^3\hat{p}}{(2\pi)^3}\hat{\Omega}_{\vec{p}}^{-2}
v^{\nu}_{\rm lat}W(x,\vec{v}_{\rm lat})\;.
\label{indcurW}\ee
with the dimensionless quantities 
\(\hat{p}_i=ap_i,\;\hat{\Omega}_{\vec{p}}=a\Omega_{\vec{p}}\) and the 
integration restricted to \(|\hat{p}_i|<\pi\).
The first term on the r.h.s. of (\ref{indcurW}) is the continuum contribution 
for which the $k$-integration decouples, as in (\ref{indcurrentW}), and has 
been performed.
In the second term on the r.h.s. of (\ref{indcurW}) the integration cannot be 
simplified since the velocity not only depends on the direction of the 
momentum \(\vec{p}\), but also on its magnitude.
For the calculation of the continuum contribution to the induced current
a field \(W(x,\vec{v})\) depending 
on the direction of \(\vec{v}\) only is sufficient, 
however the lattice contribution requires fields
that depend also on the magnitude of the velocity
\(|\vec{v}_{\rm lat}|< 1\). 
In section \ref{secfourb}
we will study the question if, for the calculation
of the Chern-Simons diffusion rate, 
we may approximate the induced current with fields that 
only depend on the direction of the velocity.

We note that the induced current (\ref{indcurW}) is covariantly conserved 
\(\left[D_{\mu},j^{\mu}_{\rm ind}\right]=0\).

Just as the usual HTL equations, the equations (\ref{claseq}), (\ref{eqW})
and (\ref{indcurW}) (or equivalently (\ref{claseq}), (\ref{eqnhtl}), 
(\ref{eqnct}) and (\ref{latindcur})) define a perturbation theory.
Taking retarded initial conditions the retarded propagator (and higher-order 
retarded vertex functions) can be obtained, as in \cite{blaizot}. 
The classical KMS 
condition then fixes the entire propagator, including its thermal part
\cite{aarts2}. 
Using perturbation 
theory we may verify
that also the time-dependent counterterms are correct, we calculate 
the retarded propagator to one-loop order. In a general gauge it 
takes the form
\be
D^{\mu\nu}_{\rm cl}(Q)=\left[g^{\mu\nu}Q^2-Q^{\mu}Q^{\nu}+F^{\mu}F^{\nu}+
\Pi_{\rm cl}^{\mu\nu}(Q)+\delta\Pi^{\mu\nu}_{\rm ind}(Q)\right]^{-1}\;,
\label{propagator}\ee
with \(F^{\mu}\) the gauge fixing vector and  \(\Pi_{\rm cl}^{\mu\nu}\)
the classical self-energy and \(\delta\Pi^{\mu\nu}_{\rm ct}\) the counterterm 
self-energy introduced in the induced source (\ref{indcurW}).
The classical self-energy to 
one-loop order reads \cite{bms,bodekerlaine}
\be
\Pi_{\rm cl}^{\mu\nu}(Q)=2g^2Na^{-1}\int\frac{d^3\hat{p}}{(2\pi)^3}
n^{\prime}_{\rm cl}(\Omega_{\vec{p}})\left[-\delta^{\mu 0}\delta^{\nu 0}+
\frac{v^{\mu}_{\rm lat}v^{\nu}_{\rm lat}q_0}{q_0+i\epsilon-\vec{v}_{\rm lat}\cdot\vec{q}}\right]\;,
\label{classe}\ee
at this order the classical self-energy  contains no contribution 
from the induced source.
The linearized induced source
\be
j^{\mu}_{\rm ind}(x)=-\int d^4x'\delta\Pi_{\rm ind}^{\mu\nu}(x,x')
A_{\nu}(x')
\label{lincur}\ee
defines the retarded self energy 
\be
\delta\Pi_{\rm ind}^{\mu\nu}(Q)=\Pi_{HTL}^{\mu\nu}(Q)-\Pi_{\rm cl}^{\mu\nu}(Q)\;,
\label{countertermse}\ee
with the continuum HTL self-energy
\be
\Pi_{HTL}^{\mu\nu}(Q)=3\omega_{\rm pl}^2\left[-\delta^{\mu 0}\delta^{\nu 0} 
+\int\frac{d\Omega}{4\pi}\frac{v^{\mu}v^{\nu}}{q_0+i\epsilon-\vec{v}\cdot\vec{q}}\right]\;.
\label{cHTLs-e}\ee
Inserting the counterterm self-energy (\ref{countertermse}) in the propagator
(\ref{propagator}), we see that the linear divergent classical self-energy
and the linear divergent part of the counterterm self-energy cancel. The 
resulting self-energy in the propagator (\ref{propagator}) is the correct
(continuum) HTL self-energy.
Finally we note that in the static limit  the self-energy (\ref{countertermse})
reduces to the counterterm mass  (\ref{massct}), as it should.

Unfortunately the system (\ref{claseq}),(\ref{eqW}), (\ref{indcurW}) is unsuitable for 
numerical implementation \cite{moorepriv}.
This can be seen from the conserved energy of the system
\be
E=\int d^3x \;\half\left[(\vec{E}^b)^2+(\vec{B}^b)^2
+3\omega_{\rm pl}^2\int\frac{d\Omega}{4\pi}W^b(x,\vec{v})W^b(x,\vec{v})
-2g^2 NTa^{-1}\int\frac{d^3\hat{p}}{(2\pi)^3}\hat{\Omega}_{\vec{p}}^{-2}
W^b(x,\vec{v}_{\rm lat})W^b(x,\vec{v}_{\rm lat})\right]\;,
\label{energy}\ee
with \(\vec{B}\) the chromo-magnetic field and \(b\) the adjoint index.
The energy is unbounded from below and this means that the system is unstable.
In an ideal world the effect of the counterterm particle density is precisely 
compensated by the hard modes of the classical gauge fields.
however in practice the evolution of the particle density and the hard modes 
will differ, which means that after some time the (wrong) effect of the 
counterterm particle density is no longer compensated by the hard modes, and 
the fields will (exponentially) blow up.

\section{Two models with a bounded energy}
\subsection{Model with lattice dispersion relation}\label{secfoura}
The goal is to obtain a model that is defined on the lattice, 
that is stable and that 
can be used to calculate IR-sensitive real-time properties of 
a non-abelian plasma without linear divergences.
Such a model should meet the following three requirements:\\
1) In the small lattice spacing limit  the continuum HTL 
equations of motion should be obtained.\\ 
2) Counterterms for the linear divergences (on the lattice) 
should be included.\\
3) The energy must be bounded from below.\\
The model considered in the previous section failed to have bounded energy.
To obtain a model with a bounded energy one can consider 
a model where the modes inducing the finite renormalization have the same 
dispersion relation as the counterterm modes.
In this section we focus on a model 
where both the counterterm modes and the modes generating the finite 
renormalization satisfy a lattice dispersion relation. 
The other possibility of enforcing the continuum dispersion relation on  
the counterterm modes is considered in the next section.

To obtain HTL equations where the both types of modes
satisfy a lattice dispersion relation, 
we do not match to a continuum quantum theory as in the previous section, 
but to a
quantum theory on the lattice, with a (small) lattice spacing \(a_S\). 
The HTL equations (in the \(A_0=0\) gauge) may then be written as 
\bea
[D_{\mu},F^{\mu\nu}(x)]&=&j_{\rm ind}^{\nu}(x)
=2gN \int\frac{d^3 \hat{p}}{(2\pi)^3}v^{\nu}_{\rm lat}
\delta \tilde{N}(x,\hat{p})\;, \label{latHTLeq1}\\
\partial_{t}\delta \tilde{N}(x,\hat{p})
-[\vec{v}_{\rm lat}\cdot \vec{D},\delta \tilde{N}(x,\hat{p})]
&=&-g\vec{v}_{\rm lat}\cdot\vec{E}(x)\;
\partial_{\hat{\Omega}_{\hat{p}}}\tilde{N}(\hat{\Omega}_{\hat{p}})\;,
\label{latHTLeq2}\eea
with \(x=(t,\vec{x})\), where the time \(t\) is continuous and
the position \(\vec{x}\) is an element of a cubic lattice with 
(large) lattice spacing \(a_{L}\).
The dimensionless momentum \(\hat{p}\) is restricted to the first Brillouin 
zone \(|\hat{p}_i|<\pi\), the dimensionless energy
\(\hat{\Omega}_{\hat{p}}=2\sqrt{\sum_{i}\sin(\hat{p}_i/2)^2}\) 
and the velocity
\(v^{i}_{\rm lat}=\partial_{\hat{p}_i}\hat{\Omega}_{\hat{p}}\).

The lattice spacing has been scaled out of the above equations and enters only
in the equilibrium distribution function \(\tilde{N}\).
The distribution function \(\tilde{N}\) should contain a contribution that, 
after solving (\ref{latHTLeq2}),
generates the quantum HTL source 
and a contribution that generates the counterterms for the classical 
divergences.
The important step is now to allow for different lattice spacings 
\(a_{L},a_{S}\)
in the the different parts of the equilibrium distribution function
\be
\tilde{N}(\hat{\Omega}_{\hat{p}})=a_{S}^{-2}n_{BE}^{S}(\hat{\Omega}_{\hat{p}})
-a_{L}^{-2}n_{\rm cl}^{L}(\hat{\Omega}_{\hat{p}})\;,
\ee
with
\bea
n_{BE}^{S}(\hat{\Omega}_{\hat{p}})&=&
\frac{1}{e^{\hat{\Omega}_{\hat{p}}/(a_S T)}-1}\;,\nn\\
n_{\rm cl}^{L}(\hat{\Omega}_{\hat{p}})&=&\frac{Ta_L}{\hat{\Omega}_{\hat{p}}}\;,
\eea
with \(T\) the temperature of the system.

To see that the model (\ref{latHTLeq1}) and (\ref{latHTLeq2}) 
contains the counterterms for the linear 
divergences it is useful to introduce the field
\be
\tilde{W}(x,\hat{p})=\delta \tilde{N}(x,\hat{p})/
\left(-g\tilde{N}^{\prime}(\hat{\Omega}_{\hat{p}})\right)\;,
\ee
which satisfies the equation
\be
\partial_{t}\tilde{W}(x,\hat{p})-
\left[\vec{v}_{\rm lat}\cdot\vec{D},\tilde{W}(x,\hat{p})\right]
=\vec{v}_{\rm lat}\cdot\vec{E}(x)\;.
\ee
The source can be split into a part generating the finite quantum HTL source
and a part subtracting the linear divergent classical source 
\be
j_{\rm ind}^{\nu}=j^{\nu}_{\rm fin}-j^{\nu}_{\rm ct}\;.
\ee
In terms of the field \(\tilde{W}\) these sources read
\bea
j_{\rm fin}^{\nu}&=&2g^2 N \int\frac{d^3 p_{S}}{(2\pi)^3}v^{\nu}_{\rm lat}
n_{BE}^{\prime}(\Omega_{S})\tilde{W}(x,\vec{p}_S a_S)\;, \label{sourcefin}\\
j_{\rm ct}^{\nu}&=&2g^2 N \int\frac{d^3 p_{L}}{(2\pi)^3}v^{\nu}_{\rm lat}
n_{\rm cl}^{\prime}(\Omega_{L})\tilde{W}(x,\vec{p}_La_L)\;,\label{sourcect}
\eea
with \(\vec{p}_S= a_S^{-1}\hat{p}\), \(\Omega_{S}=a_S^{-1}\hat{\Omega}_{\hat{p}}\)
and similar for \(\vec{p}_L, \Omega_L\).
Both sources (\ref{sourcefin}) and (\ref{sourcect}) are
covariantly conserved.

Written in dimensionfull quantities we recognize 
the source \(j_{\rm ct}\) (\ref{sourcect}) as the classical HTL source
on a lattice with lattice spacing \(a_L\). 
The difference with the perturbative model of the previous section is the 
choice of the finite renormalization.
The source \(j_{\rm fin}\) (\ref{sourcefin}) is the quantum HTL source on a 
lattice with lattice spacing \(a_S\). To extract continuum results from
this model we should require \(a_S^{-1}>>T\). Also \(a_L\) cannot be too large, since the relevant field configurations for the sphaleron rate have
size \((g^2T)^{-1}\), we should at least require \(a_L^{-1}>>g^2T\).
However as B\"{o}deker \cite{bodeker} has shown modes of spatial size \((gT)^{-1}\)
give corrections of \({\cal O}(1)\), to take these corrections into account 
requires a smaller lattice spacing \(a_L^{-1}>>gT\)

To ensure the stability of the model (\ref{latHTLeq1}) and (\ref{latHTLeq2})
we demand that the energy,
\be
E=\int d^3x \;\half\left[(\vec{E}^b)^2+(\vec{B}^b)^2-
2N\int\frac{d^3\hat{p}}{(2\pi)^3}\delta \tilde{N}^b(x,\hat{p})\delta \tilde{N}^b(x,\hat{p})/\tilde{N}^{\prime}(\hat{\Omega}_{\hat{p}})\right]\;,
\label{energymod1}\ee
is bounded from below.
This leads to the requirement
\be
-\tilde{N}^{\prime}(\hat{\Omega}_{\hat{p}})>0\;.
\label{req1}\ee
For \(\hat{p}=0\), this requirement implies \(a_S<a_L\), which is in 
accordance with the general idea that the classical theory is matched to a 
quantum theory with a smaller lattice spacing.

The function \(-\tilde{N}^{\prime}(\hat{\Omega}_{\hat{p}})\), with \(a_S<a_L\),
decreases from plus infinity at \(\hat{\Omega}_{\hat{p}}=0\), to its minimum 
below zero, after which it increases and asymptotically reaches zero.
The maximum value of the dimensionless energy is 
\(\hat{\Omega}_{\hat{p}}=2\sqrt{3}\). Demanding that
\be
-\tilde{N}^{\prime}(2\sqrt{3})>0\;,
\label{req2}\ee
together with \(a_S<a_L\) is sufficient for (\ref{req1}) to hold for any 
\(\hat{p}\). In this way, we obtain a maximum value for \(a_S^{-1}\)
given the ratio \(a_L/a_S\). In table \ref{table} some results are listed.
We see that in order to obtain continuum-like HTL contributions, the ratio
\(a_L/a_S\) should be very (exponentially) large. 

Since we want \(a_{L}^{-1}>>gT\), the coupling coupling \(g\) should be 
chosen extremely small. For instance, if we fix \(a_{S}^{-1}=2.59T\), then
stability requires \(a_L/a_S\geq 100\), so \(a_{L}^{-1}\leq 2.59\; 10^{-2}T\) 
and \(g<< 2.59\; 10^{-2}\).
\begin{table}[!t]
\caption{The maximum value of \(a_{S}^{-1}/T\) given the ration \(a_L/a_S\)
from the requirement that the energy is bounded from below.}
\begin{center}
\(\begin{array}{|l|l|l|l|l|l|l|l|l|l|l|}                                       \hline
a_L/a_S               & 1.1 & 1.5 & 2  & 5 & 10 & 20 & 25 & 50 & 100 & 1000 \\ \hline
{\rm max}(a_S^{-1})/T  & 0.31 & 0.64  & 0.86 & 1.36 & 1.68 & 1.97 & 2.06 & 2.33 & 2.59 & 3.42 \\ \hline
\end{array}\)
\end{center}
\label{table}\end{table}

To complete the model we specify the average over the initial fields, that
has to be taken to calculate classical correlation functions.
The initial conditions are
\bea
\vec{E}(t_{\rm in},\vec{x})&=&\vec{E}_{\rm in}(\vec{x})\;,\nn\\
\vec{A}(t_{\rm in},\vec{x})&=&\vec{A}_{\rm in}(\vec{x})\;,\nn\\
\delta N(t_{\rm in},\vec{x},\hat{p})&=&\delta N_{\rm in}(\vec{x},\hat{p})\;.
\label{incond}
\eea
Following \cite{iancu,iancu2}, we include the auxiliary field in the average over 
initial fields
\be
Z=\int \D E_{\rm in}\D A_{\rm in}\D\delta N_{\rm in}\;\delta\left( G_{\rm in}
\right)
\exp\left(-\beta E\right)\;,
\label{inaverage}\ee
with \(G_{\rm in}=0\) Gauss' law at the initial time. We have verified that
the phase space 
measure is invariant under time evolution.

We may conclude that the model given by (\ref{latHTLeq2}),
(\ref{incond}), (\ref{inaverage})
describes a real-time quantum theory on lattice to leading order in \(g\) 
without cut-off dependence. For very small coupling the lattice spacing 
\(a_S\) of the quantum theory can be taken small compared to the inverse 
temperature and the model describes a quantum continuum theory.
It can be used to calculate IR-sensitive quantities without lattice spacing 
dependence. Consider for instance the Chern-Simons diffusion rate, which in 
the small coupling limit takes the form \cite{bodeker}
\be
\Gamma=\left[\kappa_1 \log \frac{1}{g} + \kappa_2\right]g^{10}T^4\;.
\label{rate}\ee
A calculation of \(\Gamma\) with the model given by (\ref{latHTLeq2}),
(\ref{incond}), (\ref{inaverage}) at fixed lattice spacing \(a_S\), 
gives an \(a_L\)-independent result for both coefficients \(\kappa_1\)
and \(\kappa_2\), for \(a_L^{-1}>>gT\).

\subsection{Model with a continuum dispersion relation}\label{secfourb}

The other approach that we want to investigate is
a model where we enforce the continuum dispersion 
relation on the counterterm modes. Such a model
 has the advantage that instead
of a 6+1D field \(\delta N\) a 5+1D auxiliary field 
\(W(x,\hat{v}_{\rm lat})\), that depends only 
on the direction of the velocity 
\(\hat{v}_{\rm lat}=\vec{v}_{\rm lat}/|\vec{v}_{\rm lat}|\), can be used.
The counterterms that we obtain in this model are not exact, but for 
the calculation of the Chern-Simons diffusion rate the model 
provides a reasonable approximation.
 
The model that we consider is given by the replacement 
of the induced source (\ref{indcurrentW}) by the expression
\be
j_{\rm app}^{\nu}(x)=3\omega_{\rm pl}^2
\int\frac{d\Omega}{4\pi}v^{\nu}W(x,\vec{v})-
2g^2NT a^{-1}\int\frac{d^3\hat{p}}{(2\pi)^3}\hat{\Omega}_{\vec{p}}^{-2}
|\vec{v}_{\rm lat}|\tilde{v}^{\nu}_{\rm lat}W(x,\hat{v}_{\rm lat})\;,
\label{appsource}\ee
with \(\tilde{v}^{\nu}_{\rm lat}=(1,\hat{v}_{\rm lat})\). We choose this 
expression since it reproduces the induced vector current for 
a field configuration with \(W(x,\hat{v}_{\rm lat})=W(x,\vec{v}_{\rm lat})\), 
and the vector current 
is essential in the dynamics of the soft fields.
The density is then determined by requiring current conservation
\(\left[D_{\mu},j^{\mu}_{\rm app}\right]=0\). As a consequence the 
induced density \(j^{0}_{\rm ind}\) in (\ref{indcurW})
is not correctly reproduced by the density
\(j^{0}_{\rm app}\). This can be easily understood, changing the velocity of 
the particles and requiring current conservation either the vector current or 
the density can remain unaltered. 
The expression
(\ref{appsource}) is the lattice equivalent of the approximation for the 
induced source in \cite{iancu}.

We may also write (\ref{appsource}) as
\be
j_{\rm app}^{\nu}(x)=\int\frac{d\Omega}{4\pi}m^2(\vec{v})v^{\nu}W(x,\vec{v})\;,
\ee
with the velocity dependent mass
\be
m^2(\vec{v})=3\omega_{\rm pl}^2-
2g^2NTa^{-1}\int\frac{d^3\hat{p}}{(2\pi)^3}\hat{\Omega}_{\vec{p}}^{-2}
|\vec{v}_{\rm lat}|\delta^{S}(\vec{v}-\hat{v}_{\rm lat})\;.
\label{velmass}\ee
The second term contains a linear divergence in the direction 
\(\vec{v}=(1,1,1)/\sqrt{3}\) \cite{moorepriv} and logarithmic divergences
in directions \(\vec{v}=(1,1,s)/\sqrt{2+s^2}\) with \(-1<s<1\)
(and directions related by symmetry). Therefore the mass and the energy
are not strictly positive. To obtain a bounded energy some averaging over
the direction of the velocity \(\vec{v}\) should be performed.
This can be achieved by expanding the field \(W(x,\vec{v})\) in
spherical harmonics
\be
W(x,\vec{v})=\sum_{l m}W_{lm}(x)Y_{lm}(\vec{v})\;,
\ee
and keeping a finite number terms. 
The induced source can then be written as
\be
j^{\nu}_{\rm ind}(x)=\sum_{l m} a_{lm}^{\nu}W_{lm}(x)\;,
\ee
with coefficients
\be
a_{lm}^{\nu}=\int\frac{d\Omega}{4\pi}m^2(\vec{v})v^{\nu}
Y_{lm}(\vec{v})\;.
\ee
Given the lattice spacing \(a\), the requirement that the energy is bounded from below
puts an upper bound 
\(l_{\rm max}\) on allowed values of \(l\). It was found in \cite{bmr} 
that the 
Chern-Simons diffusion rate is insensitive to \(l_{\rm max}\) for 
even \(l_{\rm max}\). In the following we will therefore 
focus on the approximation made in (\ref{appsource}). 

As was already mentioned the approximation (\ref{appsource}) changes the 
charge density. For instance for the coefficient \(a_{00}^{0}\), we have
\be
a_{00}^{0}=m_{D}^2-2g^2NTa^{-1}\int\frac{d^3\hat{p}}{(2\pi)^3}
\hat{\Omega}_{\vec{p}}^{-2}|\vec{v}_{\rm lat}|\;.
\label{debyemassapp}\ee
Comparing 
(\ref{debyemassapp}) with (\ref{classmass}), we see that the expression
(\ref{appsource}) does not correctly reproduce the 
counterterm for the Debye mass. This implies that the current
is not suitable 
to describe the behaviour of fields at length scale \((gT)^{-1}\).

To see whether the approximation (\ref{appsource}) is valid for  fields at the
length scale \((g^2T)^{-1}\), we consider the spatial components of the 
counterterm self-energy generated by the source (\ref{appsource})
(for \(l_{\rm max}\rightarrow \infty\))
\be
\Pi_{\rm ct}^{ij}(q_0,\vec{q})=\Pi^{ij}_{HTL}(q_0,\vec{q})-
\Pi_{\rm app}^{ij}(q_0,\vec{q})\;,
\ee
with 
\be
\Pi_{\rm app}^{ij}(q_0,\vec{q})=
2g^2NTa^{-1}\int\frac{d^3\hat{p}}{(2\pi)^3}
\hat{\Omega}_{\vec{p}}^{-2}|\vec{v}_{\rm lat}|\frac{\hat{v}^{i}_{\rm lat}\hat{v}^{j}_{\rm lat}q_0}{q_{0}+i\epsilon-\hat{v}_{\rm lat}\cdot\vec{q}}\;,
\label{apps-e}\ee
which should be compared with the classical self-energy (\ref{classe}).

It is important to realize that the
relevant fields for the Chern-Simons diffusion rate we are interested in, 
have typical momenta of order \(q_0\sim g^4T,q\sim g^2 T\). 
For the gauge fields relevant for the Chern-Simons diffusion rate 
\(q_0<<|\vec{q}|\) and neglecting \(q_0\) in the denominator of the 
counterterm (\ref{apps-e}) and the classical self-energy (\ref{classe}),
we note that these are equal and that they cancel. 
For these fields the effective theory is finite and reproduces the HTL 
contributions.

However it was realized by B\"{o}deker that interactions between semi-hard 
and soft fields give corrections to the dynamics of the soft fields that are 
not suppressed by powers of \(g\). On the contrary, even \(\log(1/g)\) enhanced
contributions arise. The counterterms in the approximated source 
(\ref{appsource}) and the classical HTL's do not cancel for the semi-hard modes
(with momenta \(q_0,q\sim gT\)), therefore the semi-hard modes are sensitive 
to the cut-off \(a^{-1}\). 

The leading log contribution arises from the IR-sensitive part of the 
contribution of the semi-hard modes with
momenta \(k_0<<k\sim \mu\) and \(\mu\sim g^2T\) an IR cut-off. For these 
momenta the approximation is correct to leading order, therefore a calculation 
of the Chern-Simons diffusion rate with approximation (\ref{appsource}) 
produces the correct leading log contribution, the coefficient 
\(\kappa_1\) in (\ref{rate}) is independent of the lattice spacing.

The \({\cal O}(1)\) correction from the semi-hard modes does depend on the 
cut-off. An estimate of the cut-off dependence can be obtained from a 
comparison of the classical HTL self-energy (\ref{classe}) with the 
counterterm (\ref{apps-e}). To be explicit, we compare the diagonal components
at zero spatial momentum
\bea
\Pi^{ii}_{\rm cl}(q_0,\vec{q}=0)&=&2g^2 NTa^{-1}\int
\frac{d^3\hat{p}}{(2\pi)^3}\hat{\Omega}_{\vec{p}}^{-2}|\vec{v}_{\rm lat}|^2=
0.26 g^2 NTa^{-1}\;,\\
\Pi^{ii}_{\rm app}(q_0,\vec{q}=0)&=&2g^2 NTa^{-1}\int
\frac{d^3\hat{p}}{(2\pi)^3}\hat{\Omega}_{\vec{p}}^{-2}|\vec{v}_{\rm lat}|=
0.34 g^2NTa^{-1}\;.
\eea
Comparing the difference  between (45) and (46) with the HTL self-energy 
at zero spatial momentum
\(\Pi_{HTL}^{ii}(q_0,\vec{q}=0)=3\omega_{\rm pl}^2=g^2 T^2/3\), we obtain 
an estimate for the maximal error of about \(25\)\% for \(a^{-1}=T/\hbar\).
However, the semi-hard modes that give the \({\cal O}(1)\) correction
have space-like momenta \(q_0<|\vec{q}|\) \cite{bodeker2}. For these modes we 
expect (\ref{apps-e}) to be a better approximation 
of the classical self-energy (\ref{classe}).  

Besides the mismatch between classical HTL's and the counterterms from 
(\ref{appsource}), the lattice spacing dependence of \(\kappa_2\)
depends on the magnitude of the \({\cal O}(1)\) correction from the 
semi-hard modes. Especially when the soft modes dominate the contribution to 
\(\kappa_2\) this model is suitable for a calculation of the 
Chern-Simons diffusion rate.

\section{Conclusion}
In this paper, we studied the linear divergences in classical SU($N$) gauge 
theories at finite temperature. Counterterms for these divergences can be 
incorporated in an 
(induced) source. Although the divergences are non-local the equations of 
motion including these counterterms can be given in a local form by 
introducing auxiliary fields. In the continuum these 
auxiliary fields depend only on the direction of the velocity \(\vec{v}\), 
whereas on the lattice they necessarily also
depend on the magnitude of the velocity.
We have presented two lattice models with a bounded energy.
The first describes a real-time quantum lattice theory with a small lattice 
spacing \(a_S\). The requirement that the energy is bounded presents a lower 
bound on \(a_S\) given the lattice spacing \(a_L\) of the classical model.
It is shown that for very small coupling \(g\) the lattice spacing \(a_S\) 
can be taken small compared to the inverse temperature, which implies that 
continuum results can be obtained.

In the second model we argued that the restriction to auxiliary 
fields depending on the direction of the velocity allows for a reasonable 
approximation (\ref{appsource})
for the calculation of quantities dominated by fields with momenta 
\((q_0,q)\sim (g^4T,g^2T)\), such as the Chern-Simons diffusion rate. 

\acknowledgements
I thank Gert Aarts, Dietrich B\"{o}deker, Kari Rummukainen,
Mischa Sall\'{e}, Jan Smit and Chris van Weert for useful discussions.
I also like to thank Guy Moore for correspondence on the previous version 
of this paper.

\end{document}